\documentclass[11pt]{article}

\usepackage{amssymb,amsmath,amsfonts,amsthm,lscape}

\usepackage{graphicx}

\newcommand{\kk}{\kappa}
\newcommand{\bfI}{{\mathbb I}_\kappa}

\newcommand{\Sk}{{\rm\ \!S}}            
\newcommand{\Ck}{{\rm\ \!C}}           
 \newcommand{\Tk}{{\rm\ \!T}}           

\newcommand{\diag}{\rm diag}
\newcommand{\e}{\mathrm{e}}
\newcommand{\Id}{\rm Id}
\renewcommand{\d}{\mathrm{d}}

\begin{document}

\thispagestyle{empty}

\

 \vskip1cm

\begin{center}

\noindent {\Large{\bf {Cayley--Klein  Poisson Homogeneous Spaces
}}}

 \end{center}

\medskip
\medskip

\noindent
 {\sc Francisco J. Herranz$^\dag$\footnote{Based on the contribution presented at the
    ``XXth International Conference on Geometry, Integrability and Quantization" held in Varna, Bulgaria, June 2--7, 2018\\[4pt]
    Proceedings of the Twentieth International Conference on Geometry, Integrability and Quantization, Ivaïlo M. Mladenov, Vladimir Pulov and Akira Yoshioka, eds. (Sofia: Avangard Prima, 2019), 161-183}, Angel Ballesteros$^\dag$, Ivan Gutierrez--Sagredo$^\dag$  and\\[4pt] Mariano Santander$^\ddag$}
\medskip
\medskip

\noindent {$^\dag$Departamento de F{\'{\i}}sica, Universidad de Burgos, 
09001 Burgos, Spain 
\medskip

\noindent
$^\ddag$Departamento de F{\'{\i}}sica Te\'orica and IMUVa,  
Universidad de Valladolid,
47011 Valladolid, Spain}
 
\medskip
 
 \noindent
 {E-mails:  {\tt   fjherranz@ubu.es,  angelb@ubu.es,  igsagredo@ubu.es, \\  mariano.santander@uva.es}}

\medskip
\medskip

\begin{abstract}
\noindent
The nine two-dimensional Cayley--Klein geometries are firstly reviewed by following a graded contraction approach.
Each geometry is considered as a  set of three symmetrical homogeneous spaces (of points and two kinds of lines),   in such a manner that the graded contraction parameters determine their curvature and signature.  Secondly,  new  Poisson homogeneous spaces are constructed by making use of certain Poisson--Lie structures on the corresponding motion groups.  Therefore, the quantization of these spaces provides noncommutative analogues of the Cayley--Klein geometries. The kinematical interpretation for the  semi-Riemannian and pseudo-Riemannian Cayley--Klein geometries is   emphasized, since they are just Newtonian and  Lorentzian spacetimes of constant curvature. 
\end{abstract}

\medskip\medskip\medskip

 \medskip
 

\newpage


\section{Introduction}\label{sec:1}

The family of orthogonal Cayley--Klein (CK) algebras is a distinguished set of real  Lie  algebras that can be obtained through a graded contraction procedure
from $\mathfrak{so}(N+1)$~\cite{CKMontigny}. The CK family  depends on $N$ real contraction parameters $\kk_i$ $(i=1,\dots,N)$ and is   denoted   $\mathfrak{so}_{\kk_1,\dots,\kk_N}(N+1)$.  The relevant fact is that the CK algebra contains both semisimple and non-semsimple Lie algebras which share common geometrical and algebraical properties. The sign  of the parameters  $\kk_i$  determine a specific real form $\mathfrak{so}(p,q)$, and when (at least) one of the parameters vanishes the CK algebra becomes a non-semisimple one. Independently of the $\kk_i$ values,   all the CK algebras have the same  number of algebraically independent Casimir invariants~\cite{casimirs}, so that they have the same rank (even for the most contracted case with all $\kk_i=0$) and  they are also known as quasi-simple orthogonal algebras. From this viewpoint they can be regarded as the ``closest'' contracted algebras to the semisimple ones.
  
The
 ``Cayley--Klein'' terminology is due to the  appearance of the corresponding Lie groups ${\rm SO}_{\kk_1,\dots,\kk_N}(N+1)$ within
the context of Klein's consideration of most geometries as subgeometries
of Projective Geometry and also  to Cayley's theory of projective metrics~\cite{Ros,Yaglom,Yaglom2}. Nevertheless,  the complete
classification of these geometries was not given by Klein himself. The
two-dimensional (2D) case was studied under the name of ``quadratic geometries'' by
Poincar\'e,  following  a modern group theoretical procedure,  and the classification for arbitrary
dimension $N$ was given  by Sommerville in 1910~\cite{Sommerville}. In the latter work, he  
showed that there are $3^N$ different geometries in dimension $N$, each
corresponding to a different choice of the kind of measure of distance between
points, lines, 2-planes, \dots   which can be either elliptic, parabolic
or hyperbolic. This result can be recovered by introducing $N$ graded contraction parameters  since a positive/zero/negative value of $\kk_1,\kk_2, \kk_3,\dots$ corresponds, in this order,  to a 
kind of measure of elliptic/parabolic/hyperbolic type between points, lines, 2-planes, etc.
Furthermore, the CK groups allow for  the construction of a set of symmetrical homogeneous spaces (as coset spaces), which are interpreted as the spaces of points, lines, 2-planes, \dots, each of them of constant curvature equal to $\kk_1,\kk_2,\kk_3,\dots$~\cite{deform}. Nevertheless, in the literature only the $N$D space of points ${\rm SO}_{\kk_1,\dots,\kk_N}(N+1)/{\rm SO}_{\kk_2,\dots,\kk_N}(N)$ is usually considered.

The aim of this paper is two-fold. On the one hand, we focus on the nine 2D CK geometries and study them as a set of {\em three} symmetrical homogeneous spaces: of points and of two kinds of lines. This enables us to describe the main  properties of the CK geometries from a global approach and to explain several relations among them. On the other, we extend the notion of CK homogeneous spaces   to Poisson homogeneous spaces, which can be considered as the semiclassical counterparts of CK noncommutative spaces which are invariant under quantum deformations of the CK groups~\cite{CK2d, PL, CP, majid}.
 
The structure of the paper is as follows. In the next Section we review the nine 2D CK geometries. The   kinematical interpretation for six of them as 
Newtonian and Lorentzian spaces of constant curvature is summarized in Section 3. A set of  new ``dualities'' for the CK algebras/spaces, which generalize    the   known  ordinary duality of Projective Geometry that interchanges points with lines, is presented in Section 4. The metric structure and several sets of geodesic coordinates for the CK spaces are introduced in Section 5. Finally, we  recall the basics on  Poisson-Lie groups and quantum deformations in Section 6, which are further applied in the last Section in order to obtain  new Poisson  homogeneous spaces for the CK geometries.


\section{The Nine Two-Dimensional Cayley--Klein Geometries}\label{sec:2}

Let us consider the real Lie algebra $\mathfrak{so}(3) $ with   generators $\{J_{01},J_{02},J_{12}\}$ fulfilling the commutation rules
\begin{equation}
[J_{12},J_{01}]=J_{02},\qquad [J_{12},J_{02}]=-J_{01},\qquad [J_{01},J_{02}]=J_{12}
\nonumber
\end{equation}
and with Casimir given by
\begin{equation}
\mathcal{C}=J_{01}^2+J_{02}^2+J_{12}^2 .
\nonumber
\end{equation} 
In this basis  $\mathfrak{so}(3)$ can be endowed with a ${\mathbb Z}_2\otimes {\mathbb Z}_2$ group of commuting
involutive automorphisms generated by
\begin{equation}
\begin{aligned}
\Theta_0 &(J_{01},J_{02},J_{12})=(-J_{01},-J_{02},J_{12}) \cr
\Theta_{01} &(J_{01},J_{02},J_{12})=(J_{01},-J_{02},-J_{12})
\end{aligned}
\label{ac}
\end{equation}
such that the remaining automorphisms are the composition $\Theta_0 \Theta_{01}$
and  the identity. By applying the graded contraction theory~\cite{Montigny1,Moody}, a particular solution
of the set of  ${\mathbb Z}_2\otimes {\mathbb Z}_2$-graded contractions  from  $\mathfrak{so}(3)$  leads to a two-parametric family of Lie algebras,  
denoted as $\mathfrak{so}_{\kk_1,\kk_2}(3)$, with commutators given by~\cite{CKMontigny}
\begin{equation}
[J_{12},J_{01}]=J_{02},\qquad [J_{12},J_{02}]=-\kk_2 J_{01},\qquad [J_{01},J_{02}]=\kk_1 J_{12} 
\label{ad}
\end{equation}
where $\kk_1$ and $\kk_2$ are two real graded contraction parameters.  The corresponding Casimir reads
\begin{equation}
\mathcal{C}=\kk _2 J_{01}^2+J_{02}^2+\kk_1 J_{12}^2 .
\label{ae}
\end{equation} 
Note that each parameter $\kk_i$ $(i=1,2)$ can take any real value and it can be 
reduced to the values $\{+1,0,-1\}$ through a rescaling of the Lie algebra generators. Hence the family $\mathfrak{so}_{\kk_1,\kk_2}(3)$ comprises nine specific Lie 
algebras (some of them isomorphic). In particular,  $\mathfrak{so}_{\kk_1,\kk_2}(3)$
covers simple Lie algebras when both parameters $\kk_i\ne 0$ (the initial $\mathfrak{so}(3)$ for positive values  and 
  $\mathfrak{so}(2,1)\simeq \mathfrak{sl}_2(\mathbb R)$ otherwise),    as well as non-simple ones when at least one $\kk_i=0$  (the inhomogeneous  $
  \mathfrak{iso}(2)$,  $\mathfrak{iso}  (1,1)$ and  $\mathfrak{iiso}(1)$ where $\mathfrak{iso}(1) \equiv \mathbb {R}$). 
The relevant fact is that  $\mathfrak{so}_{\kk_1,\kk_2}(3)$   contains all the Lie algebras of the motion groups of the 2D CK geometries~\cite{CK2d,  
Gromovb, Gromova, trigo, casimirs, conf, McRae1,McRae2, Ros, Yaglom} and therefore $\mathfrak{so}_{\kk_1,\kk_2}(3)$ is called orthogonal CK algebra or quasi-simple orthogonal 
one~\cite{casimirs}.

Let us make the connection of  $\mathfrak{so}_{\kk_1,\kk_2}(3)$ with the CK geometries more explicit. 
Each automorphism (\ref{ac}) gives rise to a Cartan decomposition in the form
\begin{equation}
\begin{aligned}
\mathfrak{so}_{\kk_1,\kk_2}(3)&={\mathfrak{h}_0}\oplus  \mathfrak{p}_0,    \qquad\, 
{\mathfrak{h}_0} ={\rm span}\{ J_{12} \} ,\quad   \  \,
{\mathfrak{p}_0} ={\rm span}\{  J_{01},J_{02}\}  \cr
\mathfrak{so}_{\kk_1,\kk_2}(3)&={\mathfrak{h}_{01}}\oplus  \mathfrak{p}_{01},     \quad
{\mathfrak{h}_{01}}={\rm span}\{   J_{01} \} ,\quad 
{\mathfrak{p}_{01}} ={\rm span}\{  J_{02},J_{12}\} .
\end{aligned}
\label{af}
\end{equation}
Usually, a  2D CK geometry is understood as the \emph{set of points},  the ``plane'', which corresponds to the symmetrical homogeneous space~\cite{gil} 
coming from the first decomposition (\ref{af}) and associated with the involution $\Theta_0$.  In this way the CK homogeneous space  of points is defined by the quotient of  the
 CK Lie group  $\rm{SO}_{\kk_1,\kk_2}(3)$ by   the Lie group   $H_0$ corresponding to ${\mathfrak{h}_0}$, that is,  as the   coset space
\begin{equation}
\mathbb S^2_{[\kk_1],\kk_2}:=  {\rm  SO}_{\kk_1,\kk_2}(3)/  H_0 ,
\qquad H_0= {\rm SO}_{\kk_2}(2)=\langle J_{12}\rangle  .
\label{ag}
\end{equation}
The space $\mathbb S^2_{[\kk_1],\kk_2}$ turns out to be of constant curvature equal to $\kk_1$ and with   signature of the metric given by $\diag(+,\kk_2)$, 
so determined   by   the second parameter $\kk_2$. Therefore the generator  $J_{12}$ leaves a point $O$ invariant, the origin, generating  
 rotations around $O$. The remaining generators $J_{01}$ and $J_{02}$, that belong to the subspace $  \mathfrak{p}_0$,    generate translations which 
 move $O$ in two basic directions.

However, we can also consider the  \emph{set of lines} as  the symmetrical homogeneous space  coming from   the second decomposition (\ref{af}) and 
associated to $\Theta_{01}$, namely
\begin{equation}
\mathbb S^2_{\kk_1,[\kk_2]}:=  {\rm  SO}_{\kk_1,\kk_2}(3)/  H_{01} ,
\qquad H_{01}= {\rm SO}_{\kk_1}(2)=\langle J_{01}\rangle  .
\label{ah}
\end{equation}
The space $\mathbb S^2_{\kk_1,[\kk_2]}$ is also of constant curvature, now equal to $\kk_2$ and with   signature of the metric given by $\diag(+,\kk_1)$.
In this case, $J_{01}$ leaves a point of the space invariant (a line), while $J_{02}$ and $J_{12}$ move it,  so the former behaves as a rotation and the latter as  
translations in $\mathbb S^2_{\kk_1,[\kk_2]}$.

Moreover, it is also possible to consider a second set of lines associated to the composition $\Theta_0\Theta_{01}\equiv \Theta_{02}$ as the one defined by
the coset space
\begin{equation}
  {\rm  SO}_{\kk_1,\kk_2}(3)/  H_{02} ,
\qquad H_{02}= {\rm SO}_{\kk_1\kk_2}(2)=\langle J_{02}\rangle  .
\label{ai}
\end{equation}

Hereafter we shall call (\ref{ah}) the space of first-kind lines and   (\ref{ai}) the space of second-kind ones. By a CK geometry we will understand the set    of 
the above three  symmetrical homogeneous spaces. We display in Table~\ref{table1} the nine 2D CK geometries along with their three isotropy subgroups.


\begin{table}[t]
\caption{The nine 2D CK geometries with their specific Lie group ${\rm SO}_{\kk_1,\kk_2}(3) $ and  isotropy subgroups of    a point  $H_{0}$ (\ref{ag}), a 
first-kind line  $H_{01}$ (\ref{ah}) and a second-kind one    $H_{02}$ (\ref{ai}),  according to the value of the pair $(\kk_1,\kk_2)$.}
 \label{table1}
 \medskip 
 \medskip\medskip
\centering
\begin{tabular}{| l |  l |  l |   } 
\hline
& &   \\[-10pt]
$\bullet$ Spherical &$\bullet$ Euclidean    &$\bullet$ Hyperbolic    \\ 
 \ \   $(+,+)$:\ \ ${\rm SO}(3)$  &  \ \     $(0,+)$:\ \ ${\rm ISO}(2)$  &   \ \     $(-,+)$:\ \ ${\rm SO}(2,1)$    \\[2pt]
\ \ $H_{0}= {\rm SO}(2) $&  \ \   $H_{0}= {\rm SO}(2) $ & \ \   $H_{0}= {\rm SO}(2) $ \\ 
 \ \   $H_{01}= {\rm SO}(2) $& \ \   $H_{01}= \mathbb{R} $ &  \ \   $H_{01}= {\rm SO}(1,1) $\\ 
 \ \   $H_{02}= {\rm SO}(2) $& \ \   $H_{02}= \mathbb{R} $ & \ \   $H_{02}= {\rm SO}(1,1) $ \\[2pt]
\hline
& &   \\[-10pt]
 $\bullet$ Co-Euclidean   &$\bullet$ Galilean    &$\bullet$ Co-Minkowskian    \\ 
\ \ (Oscillating NH) &  &\ \  (Expanding NH)  \\ 
 \ \   $(+,0)$:\ \ ${\rm ISO}(2)$  &  \ \    $(0,0)$:\ \ ${\rm IISO}(1)$ &   \ \    $(-,0)$:\ \ ${\rm ISO}(1,1)$    \\[2pt]
 \ \ $H_{0}=\mathbb{R} $& \ \ $H_{0}=\mathbb{R}  $ & \ \ $H_{0}=\mathbb{R}$ \\ 
 \ \ $H_{01}={\rm SO}(2) $& \ \ $H_{01}=\mathbb{R} $ & \ \  $H_{01}={\rm SO}(1,1) $\\ 
 \ \ $H_{02}=\mathbb{R} $& \ \ $H_{02}=\mathbb{R} $ & \ \  $H_{02}=\mathbb{R} $ \\[2pt]
\hline
& &   \\[-10pt]
$\bullet$ Co-Hyperbolic   &$\bullet$ Minkowskian   &$\bullet$ Doubly Hyperbolic        \\
\ \  (Anti-de  Sitter) &  &\ \   (De Sitter)  \\ 
  \ \    $(+,-)$:\ \ ${\rm SO}(2,1)$  &  \ \    $(0,-)$:\ \ ${\rm ISO}(1,1)$ &   \ \    $(-,-)$:\ \ ${\rm SO}(2,1) $    \\[2pt]
 \ \ $H_{0}={\rm SO}(1,1)$ & \ \  $H_{0}={\rm SO}(1,1) $ &   \ \ $H_{0}={\rm SO}(1,1)$ \\
 \ \ $H_{01}={\rm SO}(2) $& \ \  $H_{01}=\mathbb{R} $ & \ \   $H_{01}={\rm SO}(1,1) $\\ 
 \ \  $H_{02}={\rm SO}(1,1) $& \ \   $H_{02}=\mathbb{R} $ & \ \   $H_{02}={\rm SO}(2) $ \\[2pt]
\hline
\end{tabular}
 \medskip 
 \medskip\medskip

 \end{table}

 
Recall that, besides their curvature/signature role,   the coefficients $\kk_i$  determine the {\em kind of measure of separation} between
points and lines in the Klein's sense \cite{Yaglom}:

\begin{itemize}
\item
 The kind of measure of distance between two   points on a first-kind line is
elliptical/ parabolical/hyperbolical according to whether $\kk_1$ is
greater than/ equal to/lesser than zero.

\item
  Likewise for  two   points on   a second-kind line depending on  
 the product $\kk_1\kk_2$.
\item  Likewise for the    kind of measure of angle  between two lines through a point
according to $\kk_2$.
\end{itemize}
Hence in the first row of Table~\ref{table1} with $\kk_2>0$, one finds the three classical Riemannian geometries with elliptical kind of measure of angles. 
The second row with $\kk_2=0$  shows the three  semi-Riemannian geometries with parabolic kind of measure of angles. And the third row with $\kk_2<0$ 
displays the pseudo-Riemannian geometries with hyperbolic kind of measure of angles. When Table~\ref{table1}  is read by columns, one sees the spaces 
of points (\ref{ag})  with positive/zero/negative curvature and with elliptical/parabolical/hyperbolical 
 kind of measure of distance between two   points on a first-kind line.

We remark that the use of the real paramaters $\kk_i$ allows for  dealing, simultaneously,   with different real forms of Lie algebras, and that making zero a given $\kk_i$ parameter corresponds to an In\"on\"u--Wigner contraction~\cite{IW,WW}. In particular, each automorphism (\ref{ac}) determines a  contraction 
which is obtained by keeping fixed the invariant generator and multplying the two anti-invariant ones by a parameter $\varepsilon$, and next taking the limit 
$\varepsilon\to 0$, that is, 
 \begin{equation}
\begin{aligned}
\Theta_0:& \quad  J'_{01}=\varepsilon J_{01},\quad  J'_{02}=\varepsilon J_{02},\quad  J'_{12}=J_{12},\quad\ \  \varepsilon\to 0\cr
\Theta_{01}:& \quad  J'_{01}=J_{01},\quad\  \, J'_{02}=\varepsilon J_{02},\quad  J'_{12}=\varepsilon J_{12},\quad  \varepsilon\to 0 
\end{aligned}
\label{aj}
\end{equation}
where $J'_{ij}$ are  the new generators.
Thus the first contraction is a local contraction, around a point, and corresponds to set $\kk_1=0$ (middle column in Table~\ref{table1}), while the second 
one is an axial contraction, around a first-kind line,  corresponding to take $\kk_2=0$ (middle row in Table~\ref{table1}).
 
It is also worth mentioning that the 2D CK geometries have been widely studied in terms of hypercomplex numbers~\cite{Gromovb, 
Gromova, Ros, Yaglom} instead of graded contraction parameters $\kk_i$. For  a detailed  use of  hypercomplex numbers
applied to the geometries with isometry group isomorphic to ${\rm SL}_2(\mathbb R)$ together with a deep insight into their properties, including   their  contractions, 
see~\cite{KisilBook, KisilProcs} and references therein.

Explicitly,  consider   real coordinates $(x,y)$ and a  hypercomplex unit $\iota$ such that
\begin{equation}
\iota^2 \in\{ -1,+1,0\}    .
\label{iota}
\end{equation}
The hypercomplex number  $z$ is defined as $ z:= x+{\iota} y $  
with conjugate    $\bar z\equiv x-\iota y$ so that
\begin{equation}
 |z|^2\equiv  z\bar z= x^2-\iota^2 y^2 .
\nonumber
\end{equation}
According to   each specific hypercomplex unit (\ref{iota})  we find the following three algebra structures on $\mathbb{R}^2$ over $\mathbb{R}$:
 \begin{itemize}
 \item If $\iota^2=-1$, then $\iota$ is an {\em elliptical}   unit leading to the usual {\em complex numbers} such that
$|z|^2= z\bar z= x^2+ y^2$.
 
\item When $\iota^2=+1$,  $\iota$ is a {\em hyperbolic}   unit providing the  so-called {\em split complex,  double or Clifford numbers}  with  $|z|^2= z\bar z= 
x^2- y^2$.   
  
\item And if $\iota^2=0$,  $\iota$ is a {\em parabolic}   unit and $z$ is known as a {\em dual or Study number},   which can be regarded as a {\em 
contracted} case since $  |z|^2= z\bar z= x^2 $. 

 \end{itemize}

From this approach,  it is necessary to consider two  hypercomplex units $\iota_1$ and $\iota_2$ to describe the 2D CK geometries, whose different possibilities  
 lead to  the nine particular geometries (see e.g.~\cite{Yaglom}), enabling one to also deal with different real forms of Lie algebras. Since the real graded 
 contraction parameters $\kk_i$  can be reduced to the standard values $\{+1,0, -1\} $, it is obvious 
that  these are somewhat related with the  hypercomplex units   $\iota_i\sim \sqrt{\kk_i}$. Hence one can naively think that both procedures are related by a mere 
identification   $\iota_i\equiv \sqrt{\kk_i}$. 
Nevertheless, the main differences   between both approaches arise in the pure contracted case  corresponding to  consider the parabolic or 
dual-Study unit with  $\iota^2=0$ and to set $\kk=0$. This  can clearly be appreciated by considering, for instance, the following contraction of exponentials of a Lie 
generator  $J$:
$$
\exp( {\iota^2}J)\to 1 ,\qquad \exp( {\iota}J)\to 1+\iota J ,\qquad \exp( {\kk}J)\to 1,\qquad  \exp( {\sqrt{\kk}}J)\to 1 .
$$
We remark that these kind of exponentials often appear in quantum group theory~\cite{CP, majid}, so that these two approaches could give rise to different  results (see 
e.g.~\cite{PLB20017} where  this fact appears explicitly in the contraction of $\mathfrak{so}_q(3)$ and $\mathfrak{so}_q(3,2)$).
 We stress that throughout the paper we will make use of the graded contraction approach, and a smooth and well-defined $\kk\to 0$ limit of all the expressions will be always feasible.


\section{Kinematical Cayley--Klein Spaces}\label{sec:3}

The six CK groups with $\kk_2\le 0$ are kinematical   groups, that is,    motion
groups of (1+1)D spacetimes of constant curvature~\cite{trigo,conf}, which are displayed in 
the second and third rows of Table~\ref{table1} (NH means Newton--Hooke). These spacetimes are the main cases within the classification of (3+1)D  kinematical 
Lie algebras formerly performed  in~\cite{BLL}  (see also~\cite{Figueroa-OFarrill2018, Figueroa-OFarrill2018higher,MontignyKinematical} and references therein).

 The geometrical-kinematical relationship is established  under the following   identification between the geometrical generators $J_{ij}$ and the   infinitesimal 
 generators of time translations  $P_0$, space translations   $P_1$ and boosts $K$:
 \begin{equation}
J_{01}\equiv P_0,\qquad J_{02}\equiv P_1,\qquad J_{12}\equiv K .
\label{da}
\end{equation}
Hence  the graded contraction parameters $\kk_i$  inherit physical dimensions   in such a manner  that they are related to the cosmological constant $\Lambda$ and the speed of light $c$, namely
 \begin{equation}
  \kk_1=-\Lambda,\qquad \kk_2=-1/c^2.
  \label{db}
\end{equation}
Thus the commutation rules (\ref{ad}) and Casimir (\ref{ae}) can be rewritten as
\begin{equation}
[K,P_0]=P_1,\qquad [K,P_1]=\frac 1{c^2} P_0,\qquad [P_0, P_1]=-\Lambda K 
\nonumber
\end{equation}
 \begin{equation}
\mathcal{C}=-\frac 1{c^2} P_0^2+P_1^2-\Lambda K^2 .
\nonumber
\end{equation}
The automorphisms (\ref{ac}) are identified with    the parity operation $\mathcal{P}\equiv \Theta_{01}$ and time-reversal  $\mathcal{T}\equiv \Theta_{02}
=\Theta_{0}\Theta_{01}$, so that the composition $\mathcal{P}\mathcal{T}\equiv \Theta_{0}$~\cite{BLL}. The substitutions $\kk_1 = 0$ and $\kk_2=  0$ 
correspond to the  spacetime  and speed-space contractions, respectively (see (\ref{aj})).

 The physical interpretation  of the three  homogeneous spaces within each of the six kinematical CK geometries is as follows: 

\begin{itemize}

\item The space  of points   (\ref{ag})  is just the (1+1)D {\em spacetime} and  its  
curvature $\kk_1$ is related to the universe (time) radius $\tau$   by $\kk_1=\pm
1/\tau^2$. The metric has signature given by $\diag(+,-1/c^2)$.

\item The space of first-kind lines (\ref{ah}) corresponds to
the 2D space of {\em time-like lines} with curvature   $\kk_2=-1/c^2$.   Its   metric now has signature $\diag(+,\kk_1)$.

\item The space of second-kind lines  (\ref{ai}) 
  is the 2D space of {\em
space-like lines}.
\end{itemize}

As it is shown in Table~\ref{table1},  the three Lorentzian
spacetimes of constant curvature $\kk_1=-\Lambda$  arise   for  $\kk_2<0$: Anti-de Sitter   ($\kk_1>0$), Minkowski   ($\kk_1=0$ or $\tau\to \infty$), and de Sitter   ($
\kk_1<0$).  Their non-relativistic limit  is provided by the contraction  $\kk_2=0$  $(c\to \infty)$, leading, in this order,  to the
oscillating  NH   ($\kk_1>0$),  Galilei  ($\kk_1=0$)
and expanding NH   ($\kk_1<0$).

Finally, we point out that besides the role of $(\kk_1,\kk_2)$ as contraction parameters, these   can also be  regarded as {\em classical deformation} ones~\cite{deform,  
Figueroa-OFarrill2018,  Figueroa-OFarrill2018higher}.
In particular, let us consider the Galilean geometry, which is the most contracted case with parameters   $(0,0)$. This means that both the spacetime and the space of time-like 
lines are flat. If a non-zero parameter $\kk_2=-1/c^2$ is introduced, then one   arrives at the Minkowskian geometry $(0,-)$ with a   curved (hyperbolic) space of  time-like lines, but 
keeping  a flat  spacetime. Next, curvature on the spacetime can be introduced through $\kk_1=-\Lambda$ giving rise to the (anti-)de Sitter spacetimes $(\kk_1,-)$. Likewise 
one can  proceed through other directions in the  deformation process.  The sequence of classical deformations ends with the (anti-)de Sitter geometries since their motion groups are always   
semisimple Lie groups ($\rm{SO}(2,1)$ at this dimension) and no  further  curvature (or physical constant) can be added if a motion Lie group is required. However, the deformation 
sequence can still continue in some sense  if quantum deformations of Lie algebras and groups are considered. In this way another deformation parameter, the ``quantum'' one $q=\e^z$, is 
introduced and, in some cases, the latter can be interpreted as a second fundamental relativistic invariant (besides $c$) which is related to the Planck scale, and thus giving rise to the so-called {\em 
Doubly Special Relativity} theories 
(see~\cite{Amelino-Camelia2010symmetry, FKS2004gravity, LN2003versus}
and  references therein).

 \newpage


 \section{Generalized Dualities}\label{sec:4}
 
  As we have already commented and can be seen in Table~\ref{table1}, some of the CK geometries have isomorphic Lie algebras.   According to $(\kk_1,\kk_2)$,  we find that  
$\mathfrak{iso}(2)$   appears twice for $(+,0)$ and $(0,+)$; $\mathfrak{iso}(1,1)$  also twice for $(-,0)$ and $(0,-)$; and $\mathfrak{so}(2,1)\simeq \mathfrak{sl}_2(\mathbb R)$ three 
times for   $(-,+)$,  $(+,-)$ and $(-,-)$. Differences among the corresponding geometries emerge when the three homogeneous spaces are taken into account altogether, which amounts 
  to focus on the isotropy subgroups of a point $H_0$, a first-kind line $H_{01}$ and a second-kind one $H_{02}$.

In fact, there exists an ``automorphism'' for the whole family of CK geometries that we shall name {\em ordinary duality} $\mathcal{D}_0$~\cite{trigo} which is defined by 
\begin{equation}
\mathcal{D}_0:\   (\tilde{J}_{01},\tilde{J}_{02},\tilde{J}_{12})=(-J_{12},-J_{02},-J_{01})
\label{ma}
\end{equation}
where $\tilde{J}_{ij}$ are the transformed generators. If we   compute the new commutation rules,   we find that these are again  (\ref{ad}) but now with transformed contraction parameters given by
$$
(\tilde{\kk}_1,\tilde{\kk}_2)=(\kk_2,\kk_1).
$$
This, in turn, means that $\mathcal{D}_0$ interchanges the spaces of points and first-kind lines, leaving the space of second-kind lines invariant, that is,
\begin{equation}
\mathbb S^2_{[\kk_1],\kk_2} \leftrightarrow  \mathbb S^2_{\kk_1,[\kk_2]} \, , \qquad H_0\leftrightarrow H_{01} .
\label{mb}
\end{equation}
Hence the Euclidean, Minkowskian and hyperbolic geometries are dual under  $\mathcal{D}_0$ to the co-Euclidean, co-Minkowskian and co-hyperbolic ones, respectively, meanwhile the 
three  remaining geometries (sphere, Galilean and doubly hyperbolic) remain invariant. Therefore  
 the prefix ``co-'' refers to this geometrical property~\cite{Yaglom} which  actually corresponds to the known duality in Projective Geometry. 
   
  Nevertheless, $\mathcal{D}_0$ does not explain other relationships  within the CK geometries   which should concern the space of second-kind lines and so explaining  
the connections   among the three geometries coming from $\mathfrak{so}(2,1)$. With this aim in mind, let us formulate the map (\ref{ma}) in terms of a permutation on the set $\mathcal S$ of indices of the generators $J_{ij}$, that is,  $\mathcal S=\{0,1,2\}$. Then  $\mathcal{D}_0$ corresponds to the 2-cycle $(0\, 2)$ and its action on the isotropy subgroups (\ref{mb})
  is consistently obtained by identifying $H_0\equiv H_{12}$. Therefore the number $3!$ of permutations on  $\mathcal S$    provides {\em six generalized dualities}, each of them being determined by a permutation element and eventually a coefficient $\kk_i$ which means that the duality cannot be applied on the geometries with $\kk_i=0$.


  \begin{table}[h]
  \caption{Transformation of the generators $J_{ij}$ and contraction parameters $(\kk_1,\kk_2)$ for the generalized dualities on the CK algebras.}\label{table2}
  \medskip
  \medskip 
  \centering
\begin{tabular}{| c |  r|  r|  r|  r|  r|  r| } 
\hline
& & & & & &  \\[-10pt]
 & \multicolumn{1}{| c|}{ $ \mathcal{D}_0$}& \multicolumn{1}{| c|}{  $ \mathcal{D}_1$} & \multicolumn{1}{| c|}{ $ \mathcal{D}_2$} &  \multicolumn{1}{| c|}{ $ \mathcal{D}_0\mathcal{D}_1$ }  &  \multicolumn{1}{| c|}{  $ \mathcal{D}_0\mathcal{D}_2$} & \multicolumn{1}{| c|}{  $ \mathcal{D}_0^2$} \\
 & $\  (0\, 2) \ $& $\ (0\, 1)\kk_1\ $ & $\ (1\, 2)\kk_2\ $ & $(0\, 2\, 1)\kk_1$ & $(0\,1\, 2)\kk_2$  & $\ \ \Id\ \ $ \\[3pt]
\hline
& & & & & &  \\[-10pt]
$\tilde{J}_{01}$ & $-J_{12}$ &$-J_{01}\ $   &  $J_{02}\ $   &   $-J_{02}\ $ &  $J_{12}\ $ &$J_{01}\ $   \\
$ \tilde{J}_{02} $& $-J_{02}$  &$\kk_1J_{12}\ $  & $\kk_2 J_{01}\ $   & $-\kk_1 J_{12}\ $ &  $-\kk_2J_{01}\ $  & $J_{02}\ $ \\
$\tilde{J}_{12} $& $-J_{01}$  & $J_{02}\ $  & $-J_{12}\ $  &  $J_{01}\ $ &  $-J_{02}\ $  & $J_{12}\ $ \\[2pt]
\hline
& & & & & &  \\[-11pt]
$\tilde{\kk}_{1}$&$\kk_2\ $  &$\kk_1\ $ &$\kk_1\kk_2\ $ &$\kk_1\kk_2\ $  & $ \kk_2\ $&  $\kk_1\ $\\
$\tilde{\kk}_{2}$ &$\kk_1\ $ & $\kk_1\kk_2\ $ &$\kk_2\ $ & $\kk_1\ $ &$\kk_1\kk_2\ $ &  $\kk_2\ $ \\[2pt]
\hline
\end{tabular}
 \end{table}


  \begin{figure}[!htbp]
  \medskip\medskip\medskip\medskip \centerline{\includegraphics[width=11.2cm]{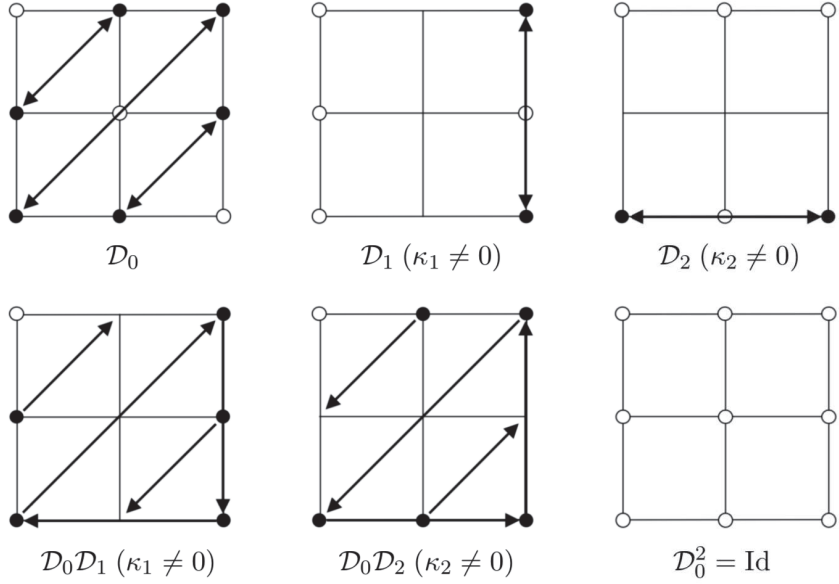}}
\caption{Generalized dualities on the nine 2D CK geometries displayed  as in Table~\ref{table1}. White circles represent invariant geometries and black ones geometries which are transformed under the  duality.}\label{figure1}
\end{figure}


  The resulting generalized dualities   are displayed in Table~\ref{table2}. Notice that only the ordinary duality (and, obviously, the identity) have no  $\kk_i$-restriction and  can be applied to the nine geometries.     We schematically represent their action on  the nine CK geometries  in Fig.~\ref{figure1}, where these   are considered in the same order (rows and columns) as in Table~\ref{table1}.
    Note also that the six dualities are always well defined on the four geometries with simple Lie group (at the corners) and  that the sphere is the only geometry which always remains invariant.

    Now let us explain some of these new results. The duality $\mathcal{D}_2=(1\, 2)\kk_2$ interchanges both spaces   of first-  and second-kind  lines, $H_{01}\leftrightarrow H_{02}$, keeping the space of points.
     The three Riemannian geometries with $\kk_2>0$ remain invariant, showing the known fact that    the sets of first- and
second-kind lines coincide (only in these cases 
the generators $J_{01}$ and $J_{02}$ are conjugated). On the three  Newtonian geometries with $\kk_2=0$,  $\mathcal{D}_2$ is not defined, which reflects that     
time-like lines are just the  ``absolute-time'' and cannot be related with the spatial lines (recall that the metric on the spacetime   is degenerate).
For the three Lorentzian  geometries  with $\kk_2<0$, this duality relates  anti-de Sitter with de Sitter, keeping Minkowskian geometry invariant. Notice that time-like lines are compact 
     in anti-de Sitter ($H_{01}={\rm SO}(2)$) while space-like lines  are non-compact ($H_{02}={\rm SO}(1,1)$) and the converse is true in de Sitter space; by contrast, in the Minkowskian case $H_{01}\equiv H_{02}\equiv \mathbb R$.

    The composition $\mathcal{D}_0\mathcal{D}_1=(0\, 2\, 1)\kk_1$, which cannot be applied to the three geometries of the second column, transforms simultaneously the three homogeneous spaces for each geometry providing the sequence
 $$
H_0\to H_{02}\to H_{01}\to H_0 .
$$
Thus the co-Euclidean geometry arrives at   the Euclidean one, but there  is no reciprocity; e.g. $H_0= {\rm SO}(2)$ for the Euclidean geometry and $H_{02}=\mathbb R$ for the co-Euclidean one   (see Table~\ref{table1}).
   Furthermore, this transformation    can be regarded as a kind of {\em ``triality''} for the three geometries associated with ${\rm SO}(2,1)$ since  
$$
 \mbox{ Hyperbolic}\  \to   \mbox{ De Sitter}\  \to\    \mbox{ Anti-de Sitter} \ \to  \mbox{ Hyperbolic}.
 $$
The two remaining dualities can be interpreted under a similar framework.


\section{Vector Model and Geodesic Coordinates for the Space of Points}\label{sec:5}

A faithful matrix representation of the CK algebra, $\rho:\mathfrak{so}_{\kk_1,\kk_2}(3)  \rightarrow \text{End} (\mathbb{R}^3)$, is given by~\cite{trigo,conf}
\begin{equation}
\rho(J_{01})=-\kk_1   e_{01}+e_{10}, \quad
\rho(J_{02})=-\kk_1\kk_2  e_{02}+e_{20}, \quad
\rho(J_{12})=-\kk_2  e_{12}+e_{21} 
\label{eb}
\end{equation}
where $e_{ij}$ is the $3\times 3$ matrix with a single non-zero  entry 1 at row $i$
and column $j$ $(i,j=0,1,2)$. It can be checked that
\begin{equation}
\rho(J_{ij})^T \bfI\,+\bfI \rho(J_{ij})=0,\qquad \bfI=\diag(1,\kk_1,\kk_1\kk_2)  .
\label{ea}
\end{equation}
Through matrix exponentiation of (\ref{eb}) we obtain  the following matrix realization
of the  isotropy subgroups of the CK group ${\rm SO}_{\kk_1,\kk_2}(3)$:
\begin{equation}
\begin{aligned}
H_{01}&={ \rm SO}_{\kk_1}(2)  :  \quad \exp (\alpha \rho(J_{01}))  =\left(\begin{array}{ccc}
\Ck_{\kk_1}(\alpha)&-\kk_1\Sk_{\kk_1}(\alpha)&0 \\[1pt] 
\Sk_{\kk_1}(\alpha)&\Ck_{\kk_1}(\alpha)&0 \\[1pt] 
0&0&1
\end{array}\right)  \\[2pt] 
H_{02}&={ \rm SO}_{\kk_1\kk_2}(2) :\  \exp(\beta \rho(J_{02}))  =\left(\begin{array}{ccc}
\Ck_{\kk_1\kk_2}(\beta)&0&-\kk_1\kk_2\Sk_{\kk_1\kk_2}(\beta) \\[1pt] 
0&1&0 \\[1pt] 
\Sk_{\kk_1\kk_2}(\beta)&0&\Ck_{\kk_1\kk_2}(\beta)
\end{array}\right)   \\[2pt] 
H_{0}&={ \rm SO}_{\kk_2}(2) : \quad \exp(\gamma \rho({J_{12} } ))=\left(\begin{array}{ccc}
1&0&0 \\[1pt] 
0&\Ck_{\kk_2}(\gamma)&-\kk_2\Sk_{\kk_2}(\gamma) \\[1pt] 
0&\Sk_{\kk_2}(\gamma)&\Ck_{\kk_2}(\gamma)
\end{array}\right) 
\end{aligned}
\label{ec}
\end{equation}
where we have introduced the $\kk$-dependent cosine  and   sine  functions~\cite{CK2d,trigo}
\begin{equation}
\Ck_{\kk}(x):=\sum_{l=0}^{\infty}(-\kk)^l\frac{x^{2l}} 
{(2l)!}=\left\{
\begin{array}{ll}
  \cos{\sqrt{\kk}\, x} &\quad  \kk>0 \\ 
\qquad 1  &\quad
  \kk=0 \\
\cosh{\sqrt{-\kk}\, x} &\quad   \kk<0 
\end{array}\right.  
\nonumber
\end{equation}
\begin{equation}
   \Sk{_\kk}(x) :=\sum_{l=0}^{\infty}(-\kk)^l\frac{x^{2l+1}}{ (2l+1)!}
= \left\{
\begin{array}{ll}
  \frac{1}{\sqrt{\kk}} \sin{\sqrt{\kk}\, x} &\quad  \kk>0 \\ 
\qquad x  &\quad
  \kk=0 \\ 
\frac{1}{\sqrt{-\kk}} \sinh{\sqrt{-\kk}\, x} &\quad  \kk<0 
\end{array}\right.  .
\nonumber
\end{equation}
The
$\kk$-tangent is defined as
$$\Tk_\kk(x)  := \frac {\Sk_\kk(x)}  { \Ck_\kk(x)}.$$
Hence these   functions are just    the  circular
and hyperbolic ones for   $\kk=\pm 1$, while under the
contraction      $\kk=0$ they reduce    to the parabolic or Galilean
functions:  $\Ck_{0}(x)=1$ and 
$\Sk_{0}(x)=\Tk_{0}(x)=x$.  Some  relations  for the above $\kk$-functions are given by~\cite{trigo}
$$
 \Ck^2_\kk(x)+\kk\Sk^2_\kk(x)=1,  \ \   \Ck_\kk(2x)= \Ck^2_\kk(x)-\kk\Sk^2_\kk(x), \ \  \Sk_\kk(2x)= 2 \Sk_\kk(x) \Ck_\kk(x) 
 $$
 and their derivatives read
 $$
  \frac{ {\rm d}}{{\rm d} x}\Ck_\kk(x)=-\kk\Sk_\kk(x),\qquad        \frac{ {\rm d}}
{{\rm d} x}\Sk_\kk(x)= \Ck_\kk(x)  ,\qquad    
\frac{ {\rm d}}
{{\rm d} x}\Tk_\kk(x)=  \frac{1}{\Ck^2_\kk(x) } \,  .
$$

 In what follows we present the metric and several sets of geodesic coordinates for the  space of points  $\mathbb S^2_{[\kk_1],\kk_2}$ (\ref{ag}).  We remark that, at this dimension, a similar procedure can be applied to the (dual)  space of first-kind lines~(\ref{ah}).

\begin{figure}
\begin{center}
 \vskip10pt
\begin{picture}(170,125)
\put(127,22){$\bullet$}
\put(52,33){\makebox(0,0){$\phi$}}
 \qbezier(46,25)(46,35)(40,38)
 \put(80,113){\makebox(0,0){$b_1$}}
\put(34,73){\makebox(0,0){$b_2$}}
\put(135,62){\makebox(0,0){$a_2$}}
\put(81,33){\makebox(0,0){$a_1$}}
\put(75,74){\makebox(0,0){$r$}}
\put(142,112){\vector(4,3){1}}
\put(117,94){$\bullet$}
\put(138,15){\makebox(0,0){$Q_1$}}
\put(137,99){\makebox(0,0){$Q$}}
\put(149,80){\makebox(0,0){$l'_1$}}
\put(149,119){\makebox(0,0){$l$}}
\put(22,22){$\bullet$}
\put(15,15){\makebox(0,0){$O$}}
\put(22,105){$\bullet$}
\put(15,108){\makebox(0,0){$Q_2$}}
\put(15,130){\makebox(0,0){$l_2$}}
\put(25,10){\vector(0,1){125}}
\put(0,25){\vector(1,0){170}}
\put(168,15){\makebox(0,0){$l_1$}}
\put(35,98){\line(0,1){10}}
\put(25,98){\line(1,0){10}}
\put(120,25){\line(0,1){10}}
\put(120,35){\line(1,0){9}}
\qbezier[50](25,108)(70,110)(140,90)
\qbezier[50](130,25)(125,80)(115,114)
\put(115,124){\makebox(0,0){$l'_2$}}
\qbezier(25,24)(50,50)(140,111)
\linethickness{1pt}
\put(70,25){\vector(1,0){20}}
\put(25,65){\vector(0,1){20}}
\put(15,73){\makebox(0,0){$J_{02}$}}
\qbezier(30,18)(40,28)(30,36)
\put(29,37){\vector(-1,1){1}}
\put(75,15){$J_{01}$}
\put(27,44){$J_{12}$}
\end{picture}
\end{center}
\caption{Infinitesimal generators $\{ J_{01},J_{02},J_{12}\}$ of isometries and  geodesic coordinates   $(a_1,a_2)$, $(b_1,b_2)$  and $(r,\phi)$    of a   point $Q=(s_0,s_1,s_2)$  on the  2D CK    space of points $\mathbb{S}^2_{[\kk_1],\kk_2}$.} 
\label{figure2}
\end{figure}
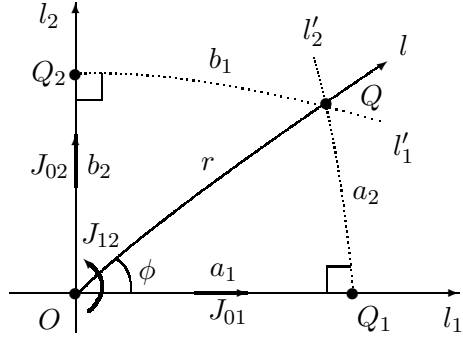

   The  matrix realization (\ref{ec})  allows us to consider the   group action of ${\rm SO}_{\kk_1,\kk_2}(3)$ on $\mathbb{R}^3$ as isometries of  the bilinear form  $\bfI$ (\ref{ea}); notice that   $  g^T \bfI\, g=\bfI $ for  a $3\times 3$ matrix  $g\in {\rm SO}_{\kk_1,\kk_2}(3) $.
Then the subgroup $H_0=  {\rm SO}_{\kk_2}(2)$  (\ref{ec})  is the isotropy subgroup of the point
$O:=(1,0,0)$, which  is thus   taken as the {\em origin} in the space
$\mathbb S^2_{[\kk_1],\kk_2}$. Therefore,   as commented in Section~\ref{sec:2}, the generator $J_{12}$ is   a  rotation on this space, while $J_{01}$ and $J_{02}$ behave as translation generators moving $O$ along two basic geodesics  $l_1$ (of first-kind)  and $l_2$ (of second-kind), which are orthogonal at $O$. This is schematically represented in Fig.~\ref{figure2}.


Next, we consider coordinates $(s_0,s_1,s_2)\in \mathbb{R}^3$.  The orbit of the origin  $O=(1,0,0)$ is contained in the ``$\kk$-sphere'' determined by $\bfI$  (\ref{ea}):
\begin{equation}
\Sigma_\kk\,  :\ s_0^2+\kk_1 s_1^2+
\kk_1\kk_2 s_2^2 =1 .
\label{ed}
\end{equation}
The connected component of $\Sigma_\kk$  can be identified with the space  ${\mathbb
S}^2_{[\kk_1],\kk_2}$ and the action of  ${\rm SO}_{\kk_1,\kk_2}(3)$ is transitive on it. 
The coordinates $(s_0,s_1,s_2)$,  satisfying  the constraint (\ref{ed})  are called {\em ambient space} or {\em Weierstrass coordinates}, 
  while $(s_1/s_0,s_2/s_0)$ are the usual {\em Beltrami coordinates}  in Projective Geometry.
  
 We define two metrics on   ${\mathbb
S}^2_{[\kk_1],\kk_2}$:   the {\em main metric} ${\rm d} \sigma_{(1)}^2$, which comes from    the flat ambient metric in $\mathbb R^{3}$ divided by the curvature $\kk_1$ and
restricted to (\ref{ed}), and a  {\em subsidiary metric} ${\rm d} \sigma_{(2)}^2$ proportional to the former one:
\begin{equation}
{\rm d} \sigma_{(1)}^2:=\left.\frac {1}{\kk_1}
\left({\rm d} s_0^2+   \kk_1 {\rm d} s_1^2+   \kk_1 \kk_2{\rm d} s_2^2  \right)\right|_{\Sigma_\kk} ,\qquad {\rm d} \sigma_{(2)}^2:= \frac 1{\kk_2}\,  {\rm d} \sigma_{(1)}^2 .
\label{ee}
\end{equation}
On the Riemannian spaces   ($\kk_2>0$) both metrics are equivalent; on the Lorentzian spacetimes  ($\kk_2<0$) these correspond, in this order, to the time- and space-like metrics; but on the Newtonian spaces with $\kk_2=0$ the main metric is degenerate with signature $(+,0)$ and there exists an invariant foliation under the action of  the CK group  ${\rm SO}_{\kk_1,0}(3)$  on ${\mathbb
S}^2_{[\kk_1], 0}$,  in such a manner that  the subsidiary metric is restricted to each leaf of the foliation.

The ambient coordinates (\ref{ed}) can be parametrized in terms of two intrinsic variables in different ways. In particular, let us introduce the so-called {\em  geodesic parallel I} $(a_1,a_2)$, {\em  geodesic parallel II} $(b_1,b_2)$   and  {\em geodesic polar} $(r,\phi)$ coordinates of a point $Q=(s_0,s_1,s_2)\in\mathbb S^2_{[\kk_1],\kk_2}$~\cite{conf}, which  are defined through the following action of the  one-parametric subgroups (\ref{ec}) on $O$:
 \begin{equation}
\begin{aligned}
(s_0,s_1,s_2)^T& = \exp(a_1\rho(J_{01})) \exp(a_2 \rho(J_{02}))O^T\cr
&=  \exp(b_2 \rho(J_{02}))\exp(b_1\rho(J_{01}))O^T\cr
&=\exp(\phi \rho(J_{12})) \exp(r \rho(J_{01}))O^T  .
\end{aligned}
\label{eff}
\end{equation}
This yields
\begin{equation}
\begin{aligned}
 s_0&=\Ck_{\kk_1}(a_1)\Ck_{\kk_1\kk_2}(a_2)=\Ck_{\kk_1}(b_1)\Ck_{\kk_1\kk_2}(b_2)=\Ck_{\kk_1}(r) \cr
  s_1&=\Sk_{\kk_1}(a_1)\Ck_{\kk_1\kk_2}(a_2)=\Sk_{\kk_1}(b_1)=\Sk_{\kk_1}(r)\Ck_{\kk_2}(\phi)  \cr
  s_2& =\Sk_{\kk_1\kk_2}(a_2)=\Ck_{\kk_1 }(b_1)\Sk_{\kk_1\kk_2}(b_2)=\Sk_{\kk_1}(r)\Sk_{\kk_2}(\phi) .
\end{aligned}
\label{ef}
\end{equation}
As shown in Fig.~\ref{figure1},   $r$ is the   distance   between the origin   $O$ and the point $Q$ measured along  the (first-kind) geodesic $l$ that joins  both points and   $\phi$ is the   angle  of $l$  with respect to  $l_1$. 
If  $Q_1$ denotes the intersection point of $l_1$ with its orthogonal (second-kind) geodesic $l_2'$ through $Q$,  then    $a_1$  is the geodesic distance between $O$ and $Q_1$ measured  along $l_1$  and    $a_2$ is the  geodesic distance   between $Q_1$ and $Q$ measured  along $l_2'$.  Similarly, for the geodesic parallel II coordinates $(b_1,b_2)$. Notice that    $(a_1,a_2) \ne (b_1,b_2)$  when $\kk_1\ne 0$.   On the flat cases   with $\kk_1=0$, the relations (\ref{ef}) show that  $s_0=1$ and
    $(s_1,s_2)=(a_1,a_2)=(b_1,b_2)$ thus reducing them to Cartesian coordinates and $(r,\phi)$ to the   polar ones. 
By introducing   (\ref{ef})   in the main  metric (\ref{ee}) we obtain  
\begin{equation}
\begin{aligned}
{\rm d} \sigma_{(1)}^2&=\Ck^2_{\kk_1\kk_2}\!(a_2){\rm d} a_1^2 + \kk_2\, {\rm d} a_2^2={\rm d} b_1^2 + \kk_2\Ck^2_{\kk_1 }\!(b_1) {\rm d} b_2^2  \cr
& =      {\rm d} r^2+\kk_2  \Sk^2_{\kk_1}\!(r)  {\rm d} \phi^2 .
\end{aligned}
\label{eg}
\end{equation}
As expected, it can easily be  checked that the Gaussian curvature is $\kk_1$ and   the signature is given by $\diag(+,\kk_2)$.

 Let us now focus  on the metric in geodesic parallel I coordinates $(a_1,a_2)$. When $\kk_2=0$, the metric is degenerate and there appears an invariant foliation determined by $a_1= a_1^0= {\rm constant}$, with subsidiary metric (\ref{ee})   $\d \sigma_{(2)}^2= \d a_2^2$  defined on each leaf  $a_1= a_1^0$. From a kinematical viewpoint (apply the identifications (\ref{da}) and (\ref{db})), it turns out that  $(a_1,a_2)$ are just the time and space variables $(x_0,x_1)$; e.g. in the Minkowskian spacetime   we recover $ {\rm d} \sigma_{(1)}^2= {\rm d} x_0^2-\frac 1 {c^2} {\rm d}x_1^2$. In the Newtonian spaces with $\kk_2=0$ $(c\to\infty)$, the main metric is just  $ {\rm d} \sigma_{(1)}^2= {\rm d} x_0^2$ which  provides ``absolute-time'' $x_0$,  the leaves of the foliation are   the ``absolute-space'' at $x_0 =x_0^0$, and  $ {\rm d} \sigma_{(2)}^2= {\rm d}x_1^2$ is the spatial metric defined on each leaf.

Isometry vector fields for the CK generators,  fulfilling (\ref{ad}), can be obtained  from the $3\times 3$  matrix representation (\ref{eb}).  In ambient coordinates $(s_0,s_1,s_2)$, satisfying  the constraint (\ref{ed}), these are given by
\begin{equation}
J_{01}=\kk_1 s_1  {\partial}_{s_0}   - s_0  {\partial}_{s_1} , \quad
J_{02}=\kk_1\kk_2 s_2 {\partial}_{s_0}   - s_0 {\partial}_{s_2} ,\quad
J_{12} =\kk_2 s_2 {\partial}_{s_1}   - s_1  {\partial}_{s_2}.
\nonumber
\end{equation}
They can be written in any geodesic coordinate system through (\ref{ef}).  For instance,   in terms of geodesic parallel I coordinates   they turn out to be   
\begin{equation}
\begin{aligned}
J_{01}&=- {\partial}_{a_1}    \cr
J_{02}&=-\kk_1\kk_2\Sk_{\kk_1}(a_1)\Tk_{\kk_1\kk_2}(a_2) {\partial}_{a_1} -\Ck_{\kk_1}(a_1) {\partial}_{a_2} \cr
 J_{12}&= \kk_2\Ck_{\kk_1}(a_1)\Tk_{\kk_1\kk_2}(a_2) {\partial}_{a_1}  -\Sk_{\kk_1}(a_1) {\partial}_{a_2} .
\end{aligned}
\label{ei}\end{equation}
Under such vector fields, the Casimir (\ref{ae}) gives rise to the Laplace--Beltrami operator on  ${\mathbb
S}^2_{[\kk_1],\kk_2}$, namely
\begin{equation}
\mathcal{C}=  
\frac{\kk_2}{\Ck_{\kk_1\kk_2}^2(a_2)}\,\partial_{a_1}^2+\partial_{a_2}^2-\kk_1\kk_2 
\Tk_{\kk_1\kk_2}(a_2)\,\partial_{a_2}  
\nonumber
\end{equation}
which, in fact,       provides   the wave equation for the Lorentzian spaces~\cite{conf} (set (\ref{db}) and $(a_1,a_2)=(x_0,x_1)$); e.g.~in the Minkowski case we have
$$
\mathcal{C}\Phi(x_0,x_1)=0 \quad \Rightarrow\quad \left( -\tfrac 1 {c^2} \partial_{x_0}^2+\partial_{x_1}^2 \right)\Phi(x_0,x_1)=0 .
$$

We illustrate the above results by presenting in Table~\ref{table3} the metric and vector fields in geodesic parallel I coordinates  for each particular space of points.



\begin{landscape}

\begin{table}[htbp]
\caption{{The nine CK spaces of points   ${\mathbf
S}^2_{[\kk_1],\kk_2}$ (\ref{ag}) according to the ``normalized'' values of the contraction parameters  $\kk_i\in\{1, 0, -1\}$. For each   space   it is shown, in geodesic parallel  I coordinates $({a_1},{a_2})$ (\ref{ef}), the    metric  (\ref{eg}) and   the Lie vector fields of isometries   (\ref{ei}). When $\kk_2=0$, the second metric is defined on ${a_1}= $ constant.}}
\medskip
\medskip\medskip
\centering
\label{table3}
\begin{tabular}{ | l | l | l | }
\hline
 & & \\[-10pt]
$\bullet$ Spherical   & $\bullet$ Euclidean    & $\bullet$ Hyperbolic \\[2pt] 
$\mathbb S^2_{[+],+}={\rm SO(3)/SO(2)}$&$ \mathbb  S^2_{[0],+}={\rm ISO(2)/SO(2)}$&
$ \mathbb  S^2_{[-],+}={\rm SO(2,1)/SO(2)}$\\[4pt]
$\d \sigma_{(1)}^2=\cos^2\! a_2 \,  \d a_1^2+ \d a_2^2$&
$\d \sigma_{(1)}^2= \d a_1^2+ \d a_2^2$&
$\d \sigma_{(1)}^2=\cosh^2\!  a_2 \, \d a_1^2+ \d a_2^2$\\[2pt] 
 $ J_{01}=-{\partial}_{a_1}  $ & $   J_{01}=-{\partial}_{a_1}   $ & $  J_{01}=-{\partial}_{a_1}   $\\[1pt] 
$  J_{02}=-\sin {a_1} \tan {a_2}\, {\partial_{a_1}} -\cos {a_1}\,{\partial_{a_2}} $ & $J_{02}=- {\partial_{a_2}} $ & $ J_{02}= \sinh {a_1} \tanh {a_2}\, {\partial_{a_1}} -\cosh {a_1}\,{\partial_{a_2}} $\\[1pt] 
$ J_{12}=\cos {a_1} \tan {a_2} \, {\partial}_{a_1}  - \sin {a_1}\, {\partial}_{a_2}  $ & $ J_{12} = {a_2} \, {\partial}_{a_1}  -   {a_1}\, {\partial}_{a_2}   $ & $ J_{12}=\cosh {a_1} \tanh {a_2} \, {\partial}_{a_1}  - \sinh {a_1} \,{\partial}_{a_2}  $
\\[4pt] 
\hline
 & & \\[-10pt]
$\bullet$ Co-Euclidean/Oscillating NH    & $\bullet$ Galilean    &$\bullet$ Co-Minkowskian/Expanding NH     \\[2pt] 
$\mathbb S^2_{[+],0}={\rm ISO(2)/\mathbb R}$&$ \mathbb  S^2_{[0],0}={\rm IISO(1)/\mathbb R}$&
$\mathbb  S^2_{[-],0}={\rm ISO(1,1)/\mathbb R}$\\[4pt]
$  {\rm d} \sigma_{(1)}^2= {\rm d} a_1^2 ,\quad    {\rm d}{\sigma}_{(2)}^2= {\rm d} a_2^2 $&
$  {\rm d} \sigma_{(1)}^2= {\rm d} a_1^2 ,\    {\rm d}{\sigma}_{(2)}^2= {\rm d} a_2^2   $&
$  {\rm d} \sigma_{(1)}^2= {\rm d} a_1^2 ,\quad    {\rm d}{\sigma}_{(2)}^2= {\rm d} {a_2}^2  $\\[2pt] 
  $  J_{01}=-{\partial}_{a_1}  $ & $  J_{01}=-{\partial}_{a_1}   $ & $ J_{01}=-{\partial}_{a_1}   $\\[1pt] 
$ J_{02}=- \cos {a_1}\,{\partial_{a_2}} $ & $ J_{02} =- {\partial_{a_2}} $ & $ J_{02}= - \cosh {a_1}\,{\partial_{a_2}} $\\[1pt] 
$ J_{12}=   - \sin {a_1}\, {\partial}_{a_2}  $ & $  J_{12}=    -   {a_1}\, {\partial}_{a_2}   $ & $ J_{12}=   - \sinh {a_1} \,{\partial}_{a_2}  $
\\[4pt] 
\hline
 & & \\[-10pt]
 $\bullet$ Co-Hyperbolic/Anti-de Sitter   & $\bullet$ Minkowskian  
&$\bullet$ Doubly Hyperbolic/De Sitter   \\[2pt] 
$\mathbb S^2_{[+],-}={\rm SO(2,1)/SO(1,1)}$&$\mathbb  S^2_{[0],-}={\rm ISO(1,1)/SO(1,1)}$&
$\mathbb  S^2_{[-],-}={\rm SO(2,1)/SO(1,1)}$\\[4pt]
$\d  \sigma_{(1)}^2=\cosh^2\! a_2 \,  \d a_1^2- \d a_2^2$&
$\d  \sigma_{(1)}^2= \d a_1^2- \d a_2^2$&
$\d  \sigma_{(1)}^2=\cos^2\! a_2 \,\d a_1^2-\d a_2^2$\\[2pt]  
 $  J_{01}=-{\partial}_{a_1}  $ & $  J_{01}=-{\partial}_{a_1}   $ & $  J_{01}=-{\partial}_{a_1}   $\\[1pt] 
$  J_{02}= \sin {a_1} \tanh {a_2}\, {\partial_{a_1}} -\cos {a_1}\,{\partial_{a_2}} $ & $ J_{02} =- {\partial_{a_2}} $ & $ J_{02}=-\sinh {a_1} \tan {a_2}\, {\partial_{a_1}} -\cosh {a_1}\,{\partial_{a_2}} $\\[1pt] 
$  J_{12}=-\cos {a_1} \tanh {a_2} \, {\partial}_{a_1}  - \sin {a_1}\, {\partial}_{a_2}  $ & $ J_{12}=- {a_2} \, {\partial}_{a_1}  -   {a_1}\, {\partial}_{a_2}   $ & $  J_{12}=-\cosh {a_1} \tan {a_2} \, {\partial}_{a_1}  - \sinh {a_1} \,{\partial}_{a_2}  $
\\[4pt] 
\hline
\end{tabular} 
\end{table}

\end{landscape}


\section{Quantum Groups   and Poisson  Homogeneous Spaces}\label{sec:6}

In this Section we introduce the basic background on quantum deformations and their connection with 
Poisson--Lie groups and Poisson  homogeneous spaces.

Let us recall that {\em quantum groups} are quantizations of  Poisson--Lie (PL)  groups {\em i.e.}, quantizations of the Poisson--Hopf algebras of multiplicative Poisson structures on Lie groups~\cite{CP, Drinfeld1987icm, majid}. PL structures $(G,\Pi)$ on a simply connected Lie group $G$ are in one-to-one correspondence with Lie bialgebra structures $(\mathfrak{g},\delta)$~\cite{Drinfeld1983hamiltonian}, where    $\mathfrak{g}$ is the Lie algebra of  $G$ and $\delta$ is the  skewsymmetric cocommutator map $\delta:\mathfrak{g} \to \mathfrak{g}\wedge \mathfrak{g}$. The cocommutator  $\delta$ must
fulfil two conditions:

\noindent
 (i) $\delta$ is a 1-cocycle,  
\begin{equation}
\delta([X,Y])=[\delta(X),\,  Y\otimes 1+ 1\otimes Y] + 
[ X\otimes 1+1\otimes X,\, \delta(Y)] ,\quad \forall  X,Y\in
\mathfrak{g}.
\nonumber
\end{equation}

\noindent
 (ii) The dual map $\delta^\ast:\mathfrak{g}^\ast\wedge \mathfrak{g}^\ast \to \mathfrak{g}^\ast$ is a Lie bracket on $\mathfrak{g}^\ast$.

Each quantum group $G_z$, with quantum deformation parameter $q=\e^z$, can be associated with a PL group $G$, and  this with a unique Lie bialgebra structure $(\mathfrak{g},\delta)$.

The dual version of quantum groups are {\em quantum algebras} ${\mathcal U}_z(\mathfrak{g})$, which are Hopf algebra deformations of universal enveloping algebras ${\mathcal U}(\mathfrak{g})$, and are constructed as formal power series in   $z$
and coefficients in ${\mathcal U}(\mathfrak{g})$. The Hopf algebra structure in ${\mathcal U}_z(\mathfrak{g})$ is provided by a coassociative coproduct map $\Delta_z: {\mathcal U}_z(\mathfrak{g})\to {\mathcal U}_z(\mathfrak{g})\otimes {\mathcal U}_z(\mathfrak{g})$, which is an algebra homomorphism,  along with the   counit $\epsilon$ and antipode $\gamma$ mappings. If we write the coproduct as a formal power series in $z$, its    first-order  determines the cocommutator in the form 
\begin{equation}
\Delta_z
=\Delta_0 + z\,\Delta_1+ o[z^2]  ,\qquad \delta =z(\Delta_1-\sigma \circ \Delta_1)
  \label{mmm}
\end{equation}
where $\Delta_0(X)=X \otimes 1+1\otimes X $ and $\sigma(X\otimes Y)=Y\otimes X$.  Hence, each quantum deformation turns out to be related to a unique Lie bialgebra structure $(\mathfrak{g},\delta)$. 
Explicitly, if we consider a basis for $\mathfrak{g}$ where
\begin{equation}
[X_i,X_j]=c^k_{ij}X_k  
\label{mma}\nonumber
\end{equation}
any cocommutator $\delta$ will be of the form
\begin{equation}
\delta(X_i)=f^{jk}_i\,X_j\wedge X_k  
\label{mmb}\nonumber
\end{equation}
being $f^{jk}_i$   the structure tensor of the dual Lie algebra $\mathfrak{g}^\ast$ 
\begin{equation}
[\hat\xi^j,\hat\xi^k]=f^{jk}_i \hat\xi^i   
\label{mc}
\end{equation}
where $\langle  \hat\xi^j,X_k \rangle=\delta_k^j$. The cocycle condition for   $\delta$   implies the following compatibility equations among the structure constants $c_{ij}^k$ and $f^{jk}_i$:
$$
f^{lm}_k c^k_{ij} = f^{lk}_i c^m_{kj}+f^{km}_i c^l_{kj}
+f^{lk}_j c^m_{ik} +f^{km}_j c^l_{ik}  .  
 $$

The connection of these structures with {\em noncommutative spaces} arises when $G$ is a group of isometries of a given space. Then $X_i$ are the Lie algebra generators and $\hat\xi^j$ can be considered as a noncommutative counterpart of the local coordinates $\xi^j$ on the group. For   a  quantum algebra ${\mathcal U}_z(\mathfrak{g})$,   the cocommutator $\delta$ is non-vanishing and the commutator~\eqref{mc} among the space coordinates associated to the translation generators of the group will be in general non-zero. This is  the way in which noncommutative spaces are constructed from quantum groups. Higher-order contributions to the noncommutative space~\eqref{mc} can be obtained from higher-orders of the full quantum coproduct  $\Delta_z$. 

In many cases the cocommutator $\delta$ is  a coboundary one, which means that it is obtained through
\begin{equation}
\delta(X)=[ X \otimes 1+1\otimes X ,\,  r],\qquad 
\forall X\in \mathfrak{g}  
\label{md}
\end{equation}
where 
$r=r^{ij}\,X_i \wedge X_j$
is an {\em $r$-matrix}, which   is a solution of the modified classical Yang--Baxter equation  
\begin{equation}
[X\otimes 1\otimes 1 + 1\otimes X\otimes 1 +
1\otimes 1\otimes X,[[r,r]]\, ]=0, \qquad \forall X\in \mathfrak{g} 
\label{me} 
\end{equation}
being $[[r,r]]$   the Schouten bracket   defined as
$$
[[r,r]]:=[r_{12},r_{13}]+ [r_{12},r_{23}]+ [r_{13},r_{23}]   
$$
  where $r_{12}=r^{ij}\,X_i \otimes X_j\otimes 1, \, r_{13}=r^{ij}\,X_i \otimes 1\otimes X_j, \, r_{23}=r^{ij}\,1 \otimes X_i\otimes X_j $. Recall that $[[r,r]]=0$ is just the classical Yang--Baxter equation.  When the PL group $G$  is a coboundary one, its Poisson structure $\Pi$   is given by the   Sklyanin bracket \cite{CP}
\begin{equation}
\{f,g\}= r^{ij} \left( \nabla^L_i f \nabla^L_j g - \nabla^R_i f \nabla^R_j g \right), \qquad f,g \in   C ^\infty(G) 
\label{mf}
\end{equation}
where  $\nabla^L_i,\nabla^R_i$ are left- and right-invariant vector fields  on  $G$.

A {\em Poisson homogeneous space} (PHS) of a PL group $(G,\Pi)$ is a Poisson manifold $(M,\pi)$ endowed with a transitive group action $\rhd: G\times M\to M$ which is a Poisson map with respect to the Poisson structure on $M$ and the product $\Pi \times \pi$ of the Poisson structures on $G$ and $M$~\cite{Drinfeld1993}. In particular, let us consider a     homogeneous space $M=G/{H}$ with     isometry Lie group $G$ and     isotropy   subgroup $H$. A PHS  $(M,\pi)$ is constructed by endowing   $G$ with the PL structure $\Pi$~\eqref{mf}, and the space  $M$ 
with a Poisson bracket $\pi$  that has to be compatible with the   group action $\rhd$.
Since $G$ may admit  several PL structures $\Pi$ (i.e., $\mathfrak g$ may admit  several Lie bialgebra structures $(\mathfrak g,\delta)$), a given homogeneous space  $M=G/H$ could lead to several  non-equivalent PHSs.

Finally,  it is known~\cite{Drinfeld1993}  that each PHS is in one-to-one correspondence with a Lagrangian subalgebra of the double Lie algebra of $\mathfrak g$ associated with the cocommutator $\delta$.  This statement is equivalent to imposing the so-called {\em coisotropy condition} for the cocommutator $\delta$ with respect to the isotropy subalgebra $\mathfrak h$ (see~\cite{BMN2017homogeneous} and references therein)
\begin{equation}
\delta(\mathfrak h) \subset \mathfrak h \wedge \mathfrak g.
\label{mg}
\end{equation}
In the more restrictive case with   $\delta\left(\mathfrak{h}\right) \subset \mathfrak{h} \wedge \mathfrak{h}$,   the subgroup $H$   have a sub-Lie bialgebra structure which   implies that the PHS is constructed through an isotropy subgroup which is a Poisson subgroup with respect to $\Pi$.


\section{Cayley--Klein Poisson Homogeneous Spaces of Points}\label{sec:7}

Among the possible classical $r$-matrices for the CK algebra $\mathfrak{so}_{\kk_1,\kk_2}(3)$ (\ref{ad}), let us consider~\cite{LBC} 
\begin{equation}
r=z J_{12}\wedge J_{02}
\label{za}
\end{equation}
which is a solution of the modified classical Yang--Baxter equation  (\ref{me}). For $  \mathfrak{so}(3)$ this generates the usual Drinfel'd--Jimbo quantum deformation, and the corresponding cocommutator (\ref{md}) reads
\begin{equation}
\delta(J_{01})=0,\qquad \delta(J_{02})=z\kk_2 J_{01}\wedge J_{02},\qquad  \delta(J_{12})=z\kk_2 J_{01}\wedge J_{12} .
\label{zb}
\end{equation}
These relations are the first-order in $z$ (\ref{mmm}) of the full coproduct for  the quantum CK algebra ${\mathcal U}_z( \mathfrak{so}_{\kk_1,\kk_2}(3) )$ which can be expressed as  
\begin{equation}
\begin{aligned}
\Delta_z(J_{01})&=J_{01}\otimes 1 + 1\otimes J_{01} \cr
\Delta_z(J_{02})&=J_{02}\otimes \exp\bigl(-\tfrac z 2 \kk_2 J_{01}\bigr) + \exp\bigl(\tfrac z 2 \kk_2  J_{01}\bigr) \otimes J_{02} \cr
\Delta_z(J_{12})&=J_{12}\otimes \exp\bigl(-\tfrac z 2 \kk_2 J_{01}\bigr) + \exp\bigl(\tfrac z 2 \kk_2  J_{01}\bigr) \otimes J_{12} .
\end{aligned}
\label{zc}
\end{equation}
This   is a homomorphism map for the deformed commutation rules given by
\begin{equation}
[J_{12},J_{01}]_z=J_{02},\quad [J_{12},J_{02}]_z=-  \frac{\sinh(z\kk_2 J_{01})}{z},\quad [J_{01},J_{02}]_z=\kk_1 J_{12} 
\nonumber
\end{equation}
which under the  classical limit $z\to 0$ reduce to the Lie brackets (\ref{ad}).

Let us denote by $\hat x^{ij}$ the quantum group coordinates dual to $J_{ij}$, that is $\langle \hat x^{ij}, J_{lm}\rangle=\delta^{ij}_{lm}$. By using  (\ref{mc})  the   cocommutator (\ref{zb}) gives rise  to the commutation relations of the dual Lie algebra  that we will denote  $\mathfrak{so}^\ast_{\kk_1,\kk_2}(3)$, namely
\begin{equation}
[\hat x^{01},\hat x^{02}]= z\kk_2\, \hat x^{02},\qquad [\hat x^{02},\hat x^{12}]= 0 ,\qquad [\hat x^{01},\hat x^{12}]= z\kk_2 \,\hat x^{12} .
\label{ze}
\end{equation}

As far as the isotropy subalgebras of a point $\mathfrak{h}_0={\rm span}\{ J_{12} \}$, a first-kind line $\mathfrak{h}_{01}={\rm span}\{ J_{01} \}$ and a second-kind one $\mathfrak{h}_{02}={\rm span}\{ J_{02} \}$ are concerned, the cocommutator (\ref{zb}) shows that  the coisotropy condition (\ref{mg}) is fulfilled for all of them:
\begin{equation}
\delta(\mathfrak h_0) \subset \mathfrak h_0 \wedge \mathfrak g ,\qquad \delta(\mathfrak h_{01})=0, \qquad \delta(\mathfrak h_{02}) \subset \mathfrak h_{02} \wedge \mathfrak g
\nonumber
 \end{equation}
 for $\mathfrak g = \mathfrak{so}_{\kk_1,\kk_2}(3)$. Then the first commutator  in (\ref{ze}) can be interpreted as  the noncommutative counterpart of the space of points $\mathbb S^2_{[\kk_1],\kk_2}$  (\ref{ag}),  at first-order in the deformation parameter and quantum coordinates, since it involves the  coordinates dual to the translation generators in this  space. Likewise, the second and third commutators in  (\ref{ze})  correspond to the (first-order)
noncommutative spaces associated with the spaces  of  first-kind  (\ref{ah}) and second-kind lines  (\ref{ai}), respectively. 

Notice that the 
noncommutative space   of  first-kind  lines is, in fact, commutative  at first-order and that the quantum deformation  ${\mathcal U}_z( \mathfrak{so}_{\kk_1,\kk_2}(3) )$  is determined by the generator $J_{01}$ which remains undeformed (\ref{zc})   (and it gives a Poisson subgroup). This is a consequence of the initial classical $r$-matrix (\ref{za}) that we have considered. Therefore we shall say that this deformation is of first-kind (or time-like)  type. Different results would come out if one starts with  $r= z J_{12}\wedge J_{01}$, which would lead to a  second-kind (or space-like)  deformation, which is just the one   fully worked out in~\cite{CK2d, PL} (see also~\cite{GromovCKq}  for quantum deformations of   CK algebras in terms hypercomplex units).

We remark that  for the three Newtonian algebras with $\kk_2=0$ the quantum deformation ${\mathcal U}_z( \mathfrak{so}_{\kk_1,\kk_2}(3) )$ is the trivial one and that (\ref{ze}) provides commutative coordinates. By contrast,  for the Lorentizan algebras with $\kk_2<0$, the first relation (\ref{ze}) defines the well-known (noncommutative) kappa-Minkowski spacetime~\cite{ LRZ1995free,Maslanka1993}. Therefore, the {\em three} Lorentzian algebras share the same noncommutative spacetime but only at the first-order in $z$ and the same happens for the Riemannian cases with $\kk_2>0$. 

 Higher-order terms can be obtained either by computing the full quantum duality or by constructing the noncommutative space as the quantization of the corresponding PHS associated with the $r$-matrix that generates the deformation. 
Let us make this statement more explicit by constructing the full noncommutative   space of points.  Let $\{a_1, a_2,\xi  \}$  be the local coordinates on the CK group $\rm{SO}_{\kk_1,\kk_2}(3)$ associated, in this order,  with the Lie generators $\{J_{01}, J_{02}, J_{12}  \}$. Then we consider the following  group element  
\begin{equation}
g = \exp(a_1\rho(J_{01})) \exp(a_2 \rho(J_{02})) \exp(\xi \rho(J_{12})) 
\label{zef}
 \end{equation}
 written under the representation (\ref{eb}).   We stress that the order chosen for the product of the matrices   enables us to identify the local coordinates $(a_1,a_2)$ with the geodesic parallel I type (\ref{eff}), so we keep the same notation.   From (\ref{zef}), the left- and right-invariant vector fields  on $\rm{SO}_{\kk_1,\kk_2}(3)$ are found to be~\cite{PL}
 \begin{align}
\nabla^L _{J_{01}}& =   \frac{\Ck_{\kk_2}(\xi) }{\Ck_{\kk_1\kk_2}(a_2) }   \, \partial_{a_1} +\Sk_{\kk_2}(\xi) \partial_{a_2}  - \kk_1\Tk_{\kk_1\kk_2}(a_2)\Ck_{\kk_2}(\xi) \, \partial_\xi  \cr
\nabla^L _{J_{02}} &=  -\kk_2\, \frac{\Sk_{\kk_2}(\xi) }{\Ck_{\kk_1\kk_2}(a_2) }   \, \partial_{a_1} +\Ck_{\kk_2}(\xi) \partial_{a_2} +\kk_1\kk_2\Tk_{\kk_1\kk_2}(a_2)\Sk_{\kk_2}(\xi) \, \partial_\xi  \cr 
\nabla^L _{J_{12}} &= \partial _\xi  
\nonumber
 \end{align}
 \begin{align}
\nabla^R _{J_{01}} & = \partial _{a_1} \cr
\nabla^R _{J_{02}}& = \kk_1\kk_2 {\Sk_{\kk_1 }(a_1) }  {\Tk_{\kk_1\kk_2}(a_2)  } \partial_{a_1} +\Ck_{\kk_1}(a_1)  \partial_{a_2} -\kk_1\, \frac{\Sk_{\kk_1}(a_1) }{\Ck_{\kk_1\kk_2}(a_2)}  \, \partial_\xi  \cr
\nabla^R _{J_{12}} &=-\kk_2 {\Ck_{\kk_1 }(a_1) }  {\Tk_{\kk_1\kk_2}(a_2)  } \partial_{a_1} +\Sk_{\kk_1}(a_1)  \partial_{a_2} +\frac{\Ck_{\kk_1}(a_1) }{\Ck_{\kk_1\kk_2}(a_2)}  \, \partial_\xi \, .
\nonumber
 \end{align}
By computing the Sklyanin bracket (\ref{mf}) for the classical $r$-matrix (\ref{za}), we obtain the Poisson structure     $\Pi$   on the CK group $\rm{SO}_{\kk_1,\kk_2}(3)$ 
\begin{equation}
\begin{aligned}
 \{\xi,a_1\}&=   - z\kk_2\, \frac{\Sk_{\kk_2}(\xi)}{\Ck_{\kk_1\kk_2}(a_2) }       ,\qquad \{\xi,a_2\}=
z \, \frac{\Ck_{\kk_1\kk_2}(a_2) \Ck_{\kk_2}(\xi)-1 }{\Ck_{\kk_1\kk_2}(a_2) } \cr
\{a_1,a_2\}&=z\kk_2 {\Tk_{\kk_1\kk_2}(a_2)  }  .
\end{aligned}
\label{zf}
\end{equation}
The canonical projection of the $\Pi$-brackets to the homogeneous space with  coordinates $(a_1,  a_2)$  gives rise to the PHS   of points   which is just the last $\pi$-Poisson bracket in (\ref{zf}). Since no ordering ambiguities appear in the r.h.s.\ of the bracket $\{a_1,a_2\}$, this can directly be  quantized   yielding the noncommutive space of points
\begin{equation}
[ \hat a_1,\hat a_2]=z\kk_2 {\Tk_{\kk_1\kk_2}(\hat a_2)  }  
\label{zff}
\end{equation}
whose first-order is given by the first commutator in (\ref{ze}) under the identification  $\hat x^{01}\equiv \hat a_1$ and $\hat x^{02}\equiv  \hat a_2$.
The noncommutative space (\ref{zff}) is, as expected, different from  the one   coming from the deformation  of   second-kind  type determined by the classical $r$-matrix $r= z J_{12}\wedge J_{01}$, namely~\cite{PL}:
\begin{equation}
[ \hat a_1,\hat a_2]=z  {\Sk_{\kk_1 }(\hat a_1)  } .
\label{zfg}
\end{equation}
From a kinematical point of view, (\ref{zfg}) can be interpreted as  space-like noncommutative   spacetimes. The noncommutative Minkowski  spacetime ($\kk_1=0$) corresponds to $[ \hat x_0,\hat x_1]=z\, \hat x_0$, and it is   different from   kappa-Minkowski~\cite{LRZ1995free,Maslanka1993} which is actually contained in (\ref{zff})  as $[ \hat x_0,\hat x_1]=- z\, \frac 1 {c^2} \, \hat x_1$. The non-relativisitic limit $\kk_2=0$ of (\ref{zfg}) now  leads to non-trivial noncommutative Newtonian spacetimes.

 
\begin{table}[t]
\caption{{The Poisson structure    (\ref{zf})   on the six CK groups $\rm{SO}_{\kk_1,\kk_2}(3)$ with $\kk_2\ne 0$ and   $\kk_i\in\{1, 0, -1\}$.  The Poisson bracket $\{a_1,a_2\}$ and $\{x_0,x_1\}$ defines the PHS of points for the Riemannian and Lorentzian cases, respectively.}}
\medskip\medskip\medskip
\centering
\label{table4}
\begin{tabular}{ | l | l | l | }
\hline
 & & \\[-10pt]
$\bullet$ Spherical   $\mathbb S^2_{[+],+} $& $\bullet$ Euclidean    $ \mathbb  S^2_{[0],+} $ & $\bullet$ Hyperbolic $ \mathbb  S^2_{[-],+} $\\[4pt] 
$\{\xi,a_1\}=   - z  \frac{\sin \xi}{\cos a_2 }   $&
$\{\xi,a_1\}=   - z \,  {\sin \xi}   $&
$\{\xi,a_1\}=   - z \frac{\sin \xi}{\cosh a_2}  $\\[4pt] 
 $  \{\xi,a_2\}=
z  \frac{\cos a_2 \cos \xi-1 }{\cos a_2 }   $ & $   \{\xi,a_2\}=
z   ({ \cos \xi-1 })   $ & $ \{\xi,a_2\}=
z   \frac{\cosh a_2 \cos \xi-1 }{\cosh a_2 } $\\[4pt] 
$ \{a_1,a_2\}=z\,  {\tan a_2 }  $ & $ \{a_1,a_2\}=z\,  a_2    $ & $ \{a_1,a_2\}=z \, {\tanh a_2 }   $
\\[4pt] 
\hline
 & & \\[-10pt]
 $\bullet$ Co-Hyperbolic   & $\bullet$ Minkowskian  
&$\bullet$ Doubly Hyperbolic  \\ 
\ \  Anti-de Sitter  $\mathbb S^2_{[+],-} $& \ \   $\mathbb  S^2_{[0],-} $ &
\ \   De Sitter $\mathbb  S^2_{[-],-} $\\[4pt]
$\{\xi,x_0\}=    z  \frac{\sinh \xi}{\cosh x_1}   $&
$\{\xi,x_0\}=    z \,  {\sinh \xi}   $&
$\{\xi,x_0\}=    z  \frac{\sinh \xi}{\cos x_1}  $\\[4pt] 
 $  \{\xi,x_1\}=
z  \frac{\cosh x_1 \cosh \xi-1 }{\cosh x_1 }   $ & $   \{\xi,x_1\}=
z (  \cosh \xi-1 )\!  $ & $ \{\xi,x_1\}=
z  \frac{\cos x_1 \cosh \xi-1 }{\cos x_1 } \!$\\[4pt] 
$ \{x_0,x_1\}=-z  \, {\tanh x_1  }  $ & $ \{x_0,x_1\}=-z \,x_1    $ & $ \{x_0,x_1\}=- z\,  {\tan x_1  }   $
\\[4pt] 
\hline
\end{tabular} \medskip\medskip
\end{table}


 To summarize, we present in Table~\ref{table4} the Poisson brackets (\ref{zf}) for the six CK groups with $\kk_2\ne 0$, since for $\kk_2= 0$ all the Poisson brackets vanish.  In the Lorentzian cases with $\kk_2<0$ the geodesic parallel I coordinates $(a_1,a_2)$ are written as the spacetime ones $(x_0,x_1)$ and $\xi$ is a rapidity. The PHSs   of first- and second-kind lines can be constructed in a similar way, but this  would require  to  consider the appropriate order for the exponentiation giving rise to the matrix group element $g$ acting transitively on the chosen coordinates.

 To end with, we stress that the classical picture of orthogonal CK algebras and (Poisson/noncommutative) spaces can be generalized to higher dimensions~\cite{LBC,deform,tesis} and  to other families of semisimple Lie algebras~\cite{tesis,MS}. Nevertheless, the quantum deformation scheme is rather involved, not only because to raise the dimension requires cumbersome computations, but  mainly  due to the fact that in higher dimensions different  possible non-equivalent deformations (and, therefore, PL structures and PHSs) can be considered. In this respect,  see~\cite{ivan} and references therein for recent results on Lorentzian kinematical algebras.


\section*{Acknowledgements}

This work was partially supported by Ministerio de Ciencia, Innovaci\'on y Universidades (Spain) under grant MTM2016-79639-P (AEI/FEDER, UE) and by the Action MP1405 QSPACE from the European Cooperation in Science and Technology (COST). I.G-S. acknowledges a predoctoral grant from Junta de Castilla y Le\'on (Spain) and the European Social Fund.


\end{document}